# Self-Repairing Disk Arrays

Jehan-François Pâris

Department of Computer Science
University of Houston
Houston, TX, USA
jfparis@uh.edu

Ahmed Amer

Department of Computer Engineering
Santa Clara University
Santa Clara, CA 95050
aamer@scu.edu

Darrell D. E. Long

Department of Computer Science
University of California
Santa Cruz, CA, USA
darrell@cs.ucsc.edu

Thomas Schwarz, S. J.

Departamento de ICC
Universidad Católica del Uruguay
Montevideo, Uruguay
tschwarz@ucu.edu.uy

*Abstract*—As the prices of magnetic storage continue to decrease, the cost of replacing failed disks becomes increasingly dominated by the cost of the service call itself. We propose to eliminate these calls by building disk arrays that contain enough spare disks to operate without any human intervention during their whole lifetime. To evaluate the feasibility of this approach, we have simulated the behavior of two-dimensional disk arrays with $n$ parity disks and $n(n - 1)/2$ data disks under realistic failure and repair assumptions. Our conclusion is that having $n(n + 1)/2$ spare disks is more than enough to achieve a 99.999 percent probability of not losing data over four years. We observe that the same objectives cannot be reached with RAID level 6 organizations and would require RAID stripes that could tolerate triple disk failures.

## I. Introduction

Despite recent advances in solid-state storage technologies, magnetic disks remain the most cost-effective solution for storing large amounts of data. As a result, most of today's large data centers store the vast majority of their data on magnetic disks. One of the main issues these centers have to face is protecting their data against potential losses caused by disk failures. All extant solutions involve a combination of maintaining enough redundant information to be able to reconstitute the contents of failed drives and promptly regenerating the contents of these drives on new ones.

These solutions are not difficult to implement in installations that have trained personnel on site round-the-clock. When this is not the case, disk repairs will have to wait until a technician can service the failed disk. There are two major disadvantages to this solution. First, it introduces an additional delay, which will have a detrimental effect on the reliability of the storage system. Second, the cost of the service call is likely to exceed that of the equipment being replaced. This is even truer for installations that are far away from metropolitan areas. Batching repairs is not an option because it would have a strong negative impact on the reliability of the array.

We present another solution. Self-repairing disk arrays are disk arrays that contain enough spare disks to free users from all maintenance tasks over the expected lifetime of each array [14, 15]. Human intervention will only be needed if the observed disk failure rates significantly exceed 4 to 5 percent per year. This would be the case if the installed disks happened to belong to a bad batch. The solution would be to replace the defective array with a new one.

Three challenges are to be met to make these organizations cost-effective. These are the initial cost of the array, its performance and its long-term reliability. The results we present here show that we can build disk arrays that can achieve 99.999 percent reliability over four years with a space overhead not exceeding that of mirroring. Our model assigns equal failure rates to both active drives and spare drives, and assumes that these rates will follow a bathtub curve with high failure rates for the first eighteen months, much lower failure rates for the next eighteen months and much higher failure rates after that.

The main lesson we learned from our study is the importance of starting with a highly reliable disk array organization. Unlike conventional disk arrays, self-repairing arrays can run out of spares. While this risk can be managed, eliminating it would require a disproportionate number of spares. As a result, self-repairing disk arrays require additional redundancy to compensate for this new risk.

Our study also illustrated the benefits of repairing failed disks as fast as possible. Disk repairs in conventional fault-tolerant arrays can be delayed by many factors, among which are the lack of spares and the need to bring in a technician. During this time interval, the array remains less protected or even completely unprotected against additional disk failures or irrecoverable read errors. Self-repairing disk arrays include their own spares and are not subject to these delays.

The remainder of this paper is organized as follows. Section II reviews previous work on fault-tolerant disk arrays. Section III introduces self-repairing disk arrays and presents the results of our simulation study. Finally, Section IV has our conclusions.

## II. Previous Work

Sparing is a well-known technique for increasing the reliability of disk arrays. Adding a spare disk to an array provides the replacement disk for the first failure. Distributed sparing [24] gains performance benefits in the initial state and degrades to normal performance after the first disk failure.

RAID arrays were the first disk array organizations to utilize *erasure coding* in order to protect data against disk failures [4, 7, 16, 21]. While RAID levels 3, 4 and 5 only tolerate single disk failures, RAID level 6 organizations use ($n - 2$)-out-of-$n$ codes to protect data against double disk failures [3]. EvenOdd, Row-Diagonal Parity and the Liberation Codes are three RAID level 6 organizations that use only XOR operations to construct their parity information [2, 4, 6, 18, 19]. Huang and Xu proposed a coding scheme correcting triple failures [9].

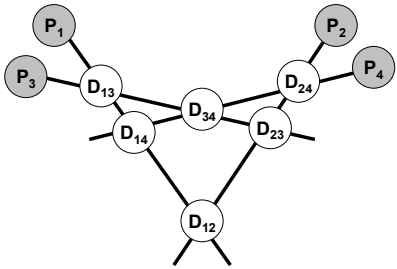

Fig. 1. A two-dimensional RAID organization using four parity disks ($P_1$ to $P_4$) to protect the contents of six data disks ($D_{12}$ to $D_{34}$) against all double disk failures.

Two-dimensional RAID arrays, or 2D-Parity arrays, were first investigated by Schwarz [22] and Hellerstein et al. [8]. More recently, Lee patented a two-dimensional disk array organization with prompt parity updates in one dimension and delayed parity updates in the second dimension [11]. Since these arrays store their parity information on dedicated disks, they are better suited for archival storage than maintaining more dynamic workloads.

Complete two-dimensional RAID arrays consist of $n$ parity stripes, each containing a single parity disk. We assume that all these stripes intersect with each other and that any two arbitrary stripes intersect at exactly one single point. We then place one data disk at each of these intersections for a total of $n(n - 1)/2$ data disks [12, 23].

Fig. 1 represents a two-dimensional RAID array with four parity disks ($P_1$ to $P_4$) and six data disks ($D_{12}$ to $D_{34}$) for a total of ten disks. More generally, a complete two-dimensional array with $n$ parity disks will comprise $n(n + 1)/2$ data disks and its space overhead will be $2/(n + 1)$.

The main interest of this organization is its high reliability. Since each data disk belongs to two distinct parity stripes, it can tolerate the failures of two arbitrary disks. As Fig. 2 shows, the sole triple failures that can result in a data loss are:

1. The failure of a data disk and its two parity disks
2. The failure of three data disks at the intersection

In addition, it has been recently shown that these arrays tolerated most quadruple and quintuple failures [23].

## III. SELF-REPAIRING DISK ARRAYS

Our main objective is to achieve 99.999 percent reliability over a five-year interval while keeping both the space overhead and the update overhead as low as possible.

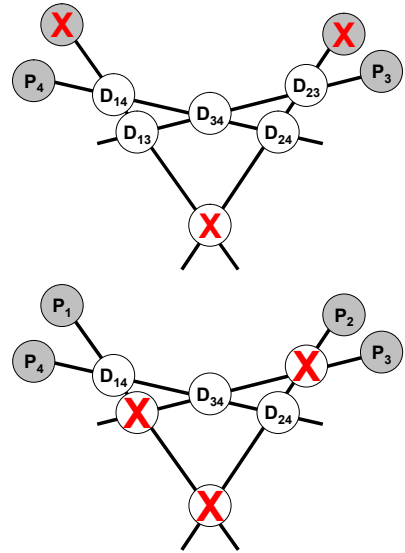

Fig. 2. Examples of fatal triple failures.

### A. Lessons from a previous attempt

In a first attempt [14], we chose three-dimensional RAID arrays [13] as our base organization because of their high reliability and their low space overhead, selected a five-year disk array lifetime and assumed disk failures were independent events distributed according to a Poisson law with a mean time to failure (MTTF) of 100,000 hours. This value is at the low end of the values observed by both Schroeder and Gibson [20], and Pinheiro, Weber and Barroso [17] and seemed to be a good conservative estimate. Disk repairs were assumed to take 12 hours.

Our simulation results indicated that the feasibility of our design depended on the failure rate of unused spare disks. As long as these rates remains negligible, zero maintenance disk arrays with at least 77 disks can provide a five-year reliability of 99.999 percent with a space overhead comparable to that of mirroring. If this is not the case, space overheads would be much higher.

We now believe these conclusions were the result of several questionable choices:

1. Assuming a disk MTTF of 100,000 hours was an overly pessimistic choice. This value corresponds to a yearly failure rate of 8.76 percent, which is much higher than the average failure rates observed by as both Schroeder and Gibson, and Pinheiro, Weber and Barroso.

2. Selecting a five-year array lifetime did not take into account that SATA drives tend to start wearing out after three years.

3. As disk capacities have increased at faster rates than their transfer bandwidth, reconstructing the contents of a failed disk on a spare will typically take more than 12 hours.

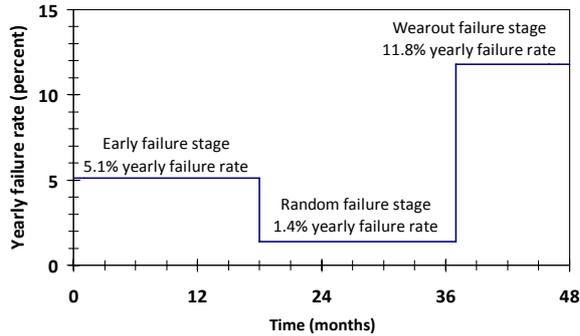

Fig. 3. Disk failure rates reported by Beach [1]

*B. Our new model*

We incorporated in our new model the conclusions of a recent study by Beach [1] reporting on the disk failure rates of more than 25,000 disks at Backblaze. He mentioned that Backblaze disks tend to fail at a rate of 5.1 percent per year during their first eighteen months, then at the much lower rate of 1.4 percent per year during the next eighteen months, and at the much higher rate of 11.8 percent per year after that. Fig. 3 summarizes these data. In our new model:

1. We assume that disks will fail at a rate of 5.1 percent per year during their first 18 months, then at 1.4 percent per year during the next 18 months, and at 11.8 percent per year after that.

2. We assume that spare disks that are yet unused fail at the same rate as the other disks, which is a very pessimistic assumption. In reality, we should expect these spare disks to fail at significantly lower than these of operational disks but have no way to estimate by how much.

3. We assume that the disk array will remain operational for four years instead of five.

4. We assume that disk repairs will take 24 hours instead of 12 hours. Assuming a disk transfer rate of 200 MB/s, this repair time suffices for rebuilding the entire contents of a 4TB disk while using less than 25 percent of the available disk bandwidth.

In addition we selected complete two-dimensional RAID arrays as our base configuration because they have lower update costs than three-dimensional RAID arrays.

*C. The simulation study*

We conducted our simulation study using the Proteus discrete simulation program [10]. Proteus characterizes each storage array by five numbers, namely the size $n$ of the array, the number $n_f$ of simultaneous disk failures the array will always tolerate without data loss, and the respective fractions $f_1$, $f_2$ and $f_3$ of simultaneous failures of $n_f + 1$, $n_f + 2$ and $n_f + 3$ disks that will not result in a data loss. The program is written in Python 3 and uses a fairly standard event-driven approach.

TABLE I. FOUR-YEAR SURVIVAL RATES AND SPACE OVERHEADS OF SELECTED TWO-DIMENSIONAL SELF-REPAIRING DISK ARRAYS.

| Number of | | | Space Overhead | Ninety-five percent C I for four-year reliability |
|---|---|---|---|---|
| Data disks | Parity disks | Spare disks | | |
| 21 | 7  | 19 | 55.32% | (4.99, 5.05) |
| 21 | 7  | 20 | 56.25% | (5.17, 5.25) |
| 28 | 8  | 23 | 52.54% | (5.00, 5.06) |
| 28 | 8  | 24 | 53.33% | (5.12, 5.20) |
| 36 | 9  | 27 | 50.00% | (4.89, 4.94) |
| 36 | 9  | 28 | 50.68% | (5.03, 5.09) |
| 45 | 10 | 33 | 48.86% | (4.98, 5.04) |
| 45 | 10 | 34 | 49.44% | (5.07, 5.13) |
| 55 | 11 | 53 | 53.78% | (4.98, 5.04) |
| 55 | 11 | 54 | 54.17% | (5.00, 5.06) |
| 66 | 12 | ∞ | ∞ | (4.79, 4.84) |

The program was slightly modified in order to handle spare disk failures differently from other disk failures and implement variable failure rates.

*C. The simulation study*

Table I summarizes our results. In the rightmost column, ninety-five percent confidence intervals for the four-year disk array reliability are expressed in "nines," using the formula $n_n = -\log_{10}(1 - R_d)$, where $n_n$ is the number of nines and $R_d$ is the four-year reliability of the array. Thus a reliability of 99.9 percent would be represented by three nines, a reliability of 99.99 percent by four nines, and so on.

As we can see, the lowest space overheads are obtained with a configuration consisting of 45 data disks, 10 parity disks and 33 or 34 spare disks. Conversely both smaller and larger array configurations require more space overhead to achieve five nine reliability over four years. In addition, the largest configuration cannot achieve five nines even with an unlimited number of spares.

To understand that, we need to understand that self-repairing arrays can fail for two different reasons. Like all fault-tolerant disk arrays, they can fail because a rapid succession of disk failures defeats the array recovery processes. In addition, they can run out of spares.

Recall that a complete two-dimensional RAID array with $n$ parity disks has $n(n - 1)/2$ data disks. Hence the parity-disk-to-data-disk ratio of any array is $2/(n - 1)$, which is a decreasing function of the size of the array. As smaller arrays have a higher parity-disk-to-data-disk ratio than larger arrays, they are inherently more reliable. Hence most data losses will occur when the array runs out of spares. The central limit theorem predicts that the coefficient of variation of the number of disk failures that will occur over a four-year interval will decrease proportionally to the square root of the size of the array. This means that guaranteeing a better than five nine probability of not running out of spares will require a much larger safety margin than for larger arrays.

Conversely, larger arrays have smaller parity-disk-to-data-disk ratios and are inherently less reliable. Obtaining five nines with a 66 disk array will thus require an ample supply of spares: Consider for instance the array configuration with 55 data disks, 11 parity disks and 54 spares in the next-to-last line of Table I. We observed 746 data losses in 80 million runs and noted that the array only ran out of spares two times during that time. Even larger arrays are inherently even less reliable. As Table I shows, an array with 66 data disks and 12 parity disks will never be able to achieve five nines over four years, even when provided with an unlimited supply of spares.

TABLE II. FOUR-YEAR SURVIVAL RATES AND SPACE OVERHEADS OF SELECTED SETS OF RAID LEVEL 6 ARRAYS WITH 12 DISKS EACH.

| Number of | | | Space Overhead | Ninety-five percent C I for four-year reliability |
|---|---|---|---|---|
| Data "disks" | Parity "disks" | Spare disks | | |
| 10 | 2 | 18 | 66.67% | (5.02, 5.09) |
| 20 | 4 | ∞ | ∞ | (4.48, 4.84) |
| 30 | 6 | ∞ | ∞ | (4.35, 4.64) |
| 40 | 8 | ∞ | ∞ | (4.33, 4.63) |

*C. Application to RAID level 6 arrays.*

We wanted to see if we could build self-repairing disk arrays by grouping together a few RAID level 6 arrays. A typical RAID level 6 array does not have separate data and parity disks. Each of its disks contains both data and parity blocks. The well-known advantage of the approach is a better update bandwidth as parity updates are spread among all the disks.

Consider a set of $m$ identical RAID level 6 arrays with $n$ disks each for a total of $m \times n$ disks. Since each RAID level 6 array will contain the equivalent of two parity disks, its parity overhead will be $2/n$. The probability that a triple disk failure will result in a data loss is the probability that all three failures occur in the same array:

$$\frac{m\binom{n}{3}}{\binom{mn}{3}}$$

The probability that a quadruple disk failure will result in a data loss is the probability that all four failures occur in the same array plus the probability that three of the four failures occur in the same array times the probability that the fourth failure occurs in one of the $m-1$ remaining arrays:

$$\frac{m\binom{n}{4} + m(m-1)n\binom{n}{3}}{\binom{mn}{4}}$$

The probability that a quintuple disk failure will result in a data loss is the probability that that all five failures occur in the same array plus the probability that four of the five failures occur in the same array times the probability that the fifth failure occurs one of the $m-1$ remaining arrays plus the probability that three of the five failures occur in the same array times the probability that the two other failures occur in the $m-1$ remaining arrays:

$$\frac{m\binom{n}{5} + m(m-1)n\binom{n}{4} + m\binom{(m-1)n}{2}\binom{n}{3}}{\binom{mn}{5}}$$

Using the same techniques as before, we simulated the performance over four year of self-repairing disk arrays consisting of one to four RAID level 6 arrays with 12 disks each. We selected that value because it happens to be a fairly popular choice for RAID level 6 due to its reasonable space overhead (2/12 = 16.7 percent).

Table II summarizes our rather disappointing results. The only configuration that could achieve five nines over four years was a single RAID array with 12 disks and it required 18 spares. Larger configurations could not achieve five nines even with an unlimited number of spares.

One obvious choice would be to lower our requirements and decide that four nines over four years would be enough. Another option would be to replace the constituting RAID level 6 arrays by arrays tolerating triple failures.

Consider, for instance, a set of $m$ identical RAID arrays tolerating triple failures. In order to be able to tolerate all triple failures, each array will have to contain the equivalent of three parity disks. The probability that a quadruple disk failure will result in a data loss is the probability that all four failures occur in the same array:

$$\frac{m\binom{n}{4}}{\binom{mn}{4}}$$

The probability that a quintuple disk failure will result in a data loss is the probability that all five failures occur in the same array plus the probability that four of the five failures occur in the same array times the probability that the fifth failure occurs in one of the $m-1$ remaining arrays:

$$\frac{m\binom{n}{5} + m(m-1)n\binom{n}{4}}{\binom{mn}{5}}$$

The probability that a sextuple disk failure will result in a data loss is the probability that that all six failures occur in the same array plus the probability that five of the six failures occur in the same array times the probability that the sixth failure occurs in one of the $m-1$ remaining arrays plus the probability that four of the six failures occur in the same array times the probability that the two other failures occur in the $m-1$ remaining arrays:

TABLE III. FOUR-YEAR SURVIVAL RATES AND SPACE OVERHEADS OF SELECTED SETS OF RAID ARRAYS WITH 15 DISKS EACH AND TRIPLE PARITY.

| Number of | | | Space Overhead | Ninety-five percent C I for four-year reliability |
|---|---|---|---|---|
| Data "disks" | Parity "disks" | Spare disks | | |
| 12 | 3 | 13 | 57.14% | (4.98, 5.17) |
| 12 | 3 | 14 | 58.62% | (5.36, 5.66) |
| 24 | 6 | 20 | 52.00% | (4.90, 5.19) |
| 36 | 9 | 26 | 49.30% | (4.98, 5.30) |
| 36 | 9 | 27 | 50.00% | (5.23, 5.97) |

$$\frac{m\binom{n}{6} + m(m-1)n\binom{n}{5} + m\binom{(m-1)n}{2}\binom{n}{4}}{\binom{mn}{6}}$$

Table III displays our results. We assumed that each RAID with triple parity consisted of 15 disks. Since each of these arrays contains the equivalent of three parity disks, its parity overhead is 3/15, that is, 20 percent. As we can see, the new organization can achieve five-nine reliability over four years with a space overhead comparable to that of mirroring. As we observed before, the very small configurations have a significantly higher space overhead than the bigger ones.

IV. CONCLUSION

We have proposed to reduce the maintenance cost of disks arrays by building self-repairing arrays that contain enough spare disks to operate without any human intervention during their whole lifetime. To illustrate the feasibility of our approach, we have shown that several complete two-dimensional disk arrays with $n$ parity disks, $n(n-1)/2$ data disks, and less than $n(n+1)/2$ data disks could achieve a 99.999 percent probability of not losing data over four years. We also noted that the same objectives cannot be reached with self-repairing arrays consisting of RAID level 6 parity stripes and would require RAID stripes capable of tolerating three disk failures.

More work is still needed to define policies that would allow array users and manufacturers to detect unusually disk failure rates and take the appropriate actions before any data loss takes place.

ACKNOWLEDGMENTS

A. A. and D. D. E. L. were supported in part by the National Science Foundation under awards CCF-1219163 and CCF-1217648, by the Department of Energy under award DE-FC02-10ER26017/DE-SC0005417, by the industrial members of the Storage Systems Research Center and by a gift from Wells Fargo.